\documentclass[a4paper]{jpconf}
\usepackage{amsmath,slashed}
\usepackage{graphicx,graphics,color}
\usepackage{dcolumn}
\usepackage[hyperfootnotes=false]{hyperref}
\usepackage{xspace}
\usepackage{cite}

\newcommand{\beq}{\begin{equation}}
	\newcommand{\eeq}{\end{equation}}

\allowdisplaybreaks[1]

\begin{document}
	\title{Beyond the Standard Model with Lepton Flavor Universality Violation}
	
	\author{Andreas Crivellin}
		\ead{andreas.crivellin@cern.ch}
\address{Physik-Institut, Universit\"at Z\"urich, Winterthurerstrasse 190, CH--8057 Z\"urich, Switzerland}
\address{Paul Scherrer Institut, CH--5232 Villigen PSI, Switzerland}
\address{CERN Theory Division, CH--1211 Geneva 23, Switzerland}
\author{Joaquim Matias}
\address{Universitat Autonoma de Barcelona and IFAE, E-08193 Bellaterra, Spain}

	\begin{abstract}
		In recent years, exciting (indirect) hints for physics beyond the Standard Model (SM) have been accumulated. In particular, semi-leptonic $B$ decays show deviations from the SM predictions, which, due to the ratios $R(K^{(*)})$ and $R(D^{(*)})$ are obviously related to lepton flavour universality violation (LFUV). However, {we point out} there are more anomalies which admit an interpretation in terms of LFUV: The anomalous magnetic moment of the muon, the Cabibbo angle anomaly, the CMS measurements of non-resonant di-electrons, the difference of the forward-backward asymmetry in $B\to D^*\ell\nu$ and leptonic tau decays. In this letter we discuss the experimental and theoretical status of these anomalies, {compare their strength and weaknesses} and examine {and synthesize} how they can be explained in terms of possible extensions of the SM by new particles and interactions. Even though not all anomalies might be confirmed in the future, this unified view of the anomalies in terms of LFUV significantly strengthens their relevance, which is crucial in order to construct a convincing physics case for future colliders.
	\end{abstract}

	\section{Introduction}
	
	The SM of particle physics has been tested and confirmed by many indirect and direct measurements in the last decades~\cite{ParticleDataGroup:2020ssz} and it was completed in 2012 by the discovery of the Brout-Englert-Higgs boson~\cite{ATLAS:2012yve,CMS:2012qbp}. Therefore, the focus of particle physics has now shifted towards discovering physics beyond the SM, i.e. new particles and new interactions. However, despite the observation of Dark Matter (at astrophysical scales) and neutrino masses (via oscillations), as well as compelling theoretical arguments for the existence of beyond the SM physics, no new particles were (so far) observed directly at the Large Hadron Collider (LHC) at CERN (see e.g. Refs.~\cite{Butler:2017afk,Masetti:2018btj} for an overview). 
	Fortunately, intriguing indirect hints for new physics (NP) have been accumulated in:
	\begin{itemize}
		\item Semi-leptonic bottom quark decays ($b\to s\ell^+\ell^-$) 
		\item Tauonic $B$ meson decays ($b\to c\tau\nu$) 
		\item The anomalous magnetic moment of the muon ($a_\mu$)
		\item	The Cabibbo angle anomaly (CAA) 
		\item	Non-resonant di-electrons ($q\bar q \to e^+e^-$)
		\item The difference of the forward-backward asymmetry in $B\to D^*\mu\nu$ vs $B\to D^*e\nu$ ($\Delta A_{\rm FB}$)
		\item Leptonic tau decays ($\tau\to\mu\nu\nu$)
	\end{itemize}
	within recent years. Interestingly, as we want to {propose} in this letter, all these observables admit an interpretation in terms of LFUV, i.e. NP that distinguishes between muon, electrons and tau leptons. While some of the anomalies are by construction measures of LFUV, also the other observables can be interpreted in this context. This unified view suggests a common origin of the anomalies in terms of beyond the SM (BSM) physics which reinforces the case for LFUV. {In particular, it opens up novel avenues for the construction of NP models and allows for the construction of a compelling physics case for colliders.}
	
	\section{Anomalies and LFUV}
	
	In the SM, the gauge interactions respect lepton flavour universality (LFU), which is in fact only broken by the Higgs Yukawa couplings. As these couplings are very small, at most of the order of one percent for the tau lepton, LFU is an approximate accidental symmetry of the SM (at the Lagrangian level). However, the impact of the lepton masses, originating from the Higgs Yukawa couplings after electroweak (EW) symmetry breaking, on the life times of charged leptons is enormous due to kinematic effects. Therefore, if we refer to LFUV we mean interactions with different couplings to electrons, muons and tau leptons (disregarding phase space effects) that directly distinguish among the charged leptons at the Lagrangian level. 
	
	\vspace{2mm}
	\begin{boldmath}$b\to s\ell^+\ell^-$:\end{boldmath}~As all flavour changing neutral current processes, $b\to s\ell^+\ell^-$ transitions are loop and CKM suppressed in the SM, resulting in branching ratios which are at most of the order of $10^{-6}$. Among the different decays involving $b\to s\ell^+\ell^-$ transitions, the ratios 
	$R({K^{\left(*\right)}})  = {\rm Br}({B \to K^{\left( * \right)}\mu ^+\mu ^-})
	/{\rm Br}({B \to K^{\left( * \right)}e^+e^-})$
	are particularly prominent. They are measured by LHCb~\cite{LHCb:2021trn,LHCb:2017avl} (and Belle~\cite{BELLE:2019xld,Belle:2019oag}) and their theory predictions are very clean (within the SM) since the dependence on the form factors drops out to an excellent approximation. Moreover, there is a long list of semi-leptonic $b \to s \ell^+\ell^-$ observables that significantly deviate from the SM predictions, like the optimized  observable $P_5^{\prime\mu}$~\cite{Descotes-Genon:2012isb,Descotes-Genon:2013vna}\footnote{$P_5^{\prime\mu}$ leads to the important LFUV observable $Q_5=P_5^{\prime\mu}-P_5^{\prime e}$~\cite{Capdevila:2016ivx} measured by Belle\cite{Belle:2016fev} which agrees with the expectations from $R(K^{(*)})$.}, but also total branching ratios (sensitive to form factors) like ${\rm Br}({B\to K^{*}\mu^+\mu^-})$~\cite{LHCb:2016ykl}, ${\rm Br}({B \to K\mu^+\mu^-})$\cite{LHCb:2014cxe} and ${\rm Br}({B_s \to \phi\mu^+\mu^-})$~\cite{LHCb:2021zwz}. Furthermore, also the decay $B_s\to\mu^+\mu^-$~\cite{LHCb:2020zud}, being a purely leptonic decay, which can be predicted accurately including 3-loop QCD corrections~\cite{Hermann:2013kca} and enhanced electromagnetic corrections~\cite{Beneke:2017vpq}, displays a tension. Even though none of these observables measures LFUV, it is intriguing that if one assumes that the bulk of the NP effect is related to muons, a picture emerges which is in excellent agreement with data, further reinforcing the case for BSM physics in $b\to s\ell^+\ell^-$ transitions. In fact, including all observables into a global fit~\cite{Descotes-Genon:2015uva,Capdevila:2017bsm,Alguero:2019ptt}, one finds that several NP scenarios, possessing destructive NP w.r.t. the SM at the $20\%$ level, are preferred over the SM hypothesis by more than $7\sigma$~\cite{Alguero:2021anc} (see also Refs.~\cite{Altmannshofer:2021qrr,Alok:2020bia,Hurth:2020ehu,
		Ciuchini:2020gvn}).
	
	\vspace{2mm}
	\begin{boldmath}$b\!\!\to\!\! c\tau\nu$:\end{boldmath} This charged current transition is already mediated at tree-level in the SM and the corresponding decays have significant branching ratios (${\mathcal O}(10^{-3})$). Here the ratios
	$R\left( {{D^{\left( * \right)}}} \right) = {{\rm Br}({B \to {D^{\left( * \right)}}\tau \nu }
		)/ {\rm Br}({ B \to {D^{\left( * \right)}}\ell \nu )})}$, measured by BaBar~\cite{BaBar:2013mob}, Belle~\cite{Belle:2019gij} and LHCb~\cite{LHCb:2017smo}, possess imperfect cancellations of the form factor dependence since the tau mass is sizeable. However, the error is experimentally dominated and the resulting significance for constructive NP, at the $10\%$ level w.r.t. the SM, is $3\sigma$~\cite{HFLAV:2019otj}. Interestingly, these measurements are supported by $R(J/\Psi)$~\cite{LHCb:2017vlu} which is also observed to be larger than expected in the SM.
	
	\vspace{2mm}
	\begin{boldmath}$a_\mu$:\end{boldmath}~The result of the E821 experiment at Brookhaven~\cite{Bennett:2006fi} was recently confirmed by the $g-2$ experiment at Fermilab~\cite{Abi:2021gix} and the combined result displays a $4.2\,\sigma$ tension with the SM prediction~\cite{Aoyama:2012wk,Aoyama:2019ryr,Czarnecki:2002nt,Gnendiger:2013pva,Davier:2017zfy,Keshavarzi:2018mgv,Colangelo:2018mtw,Hoferichter:2019gzf,Davier:2019can,Keshavarzi:2019abf,Kurz:2014wya,Melnikov:2003xd,Masjuan:2017tvw,Colangelo:2017fiz,Hoferichter:2018kwz,Gerardin:2019vio,Bijnens:2019ghy,Colangelo:2019uex,Blum:2019ugy,Colangelo:2014qya}, which points towards NP of the order of the SM EW contribution. The main theory uncertainty comes from hadronic vacuum polarization (HVP). Here, the Budapest Marseilles Wuppertal collaboration (BMWc) released first lattice results that indicate a value in tension with the one derived from $e^+e^-$ data~\cite{Davier:2017zfy,Keshavarzi:2018mgv,Colangelo:2018mtw,Hoferichter:2019gzf,Davier:2019can,Keshavarzi:2019abf}. Therefore, future detailed comparisons with other lattice calculations will be necessary to achieve the same level of scrutiny that has become standard for the data-driven approach\footnote{Note that HVP also enters the global EW fit~\cite{Passera:2008jk}, and its {indirect determination via this fit~\cite{Haller:2018nnx}, slightly} disfavours the BMWc result~\cite{Crivellin:2020zul,Keshavarzi:2020bfy}.}.
	Note that {the UV contribution to} $a_\ell$ is necessarily proportional to one power of the lepton mass due to kinematics. Factoring out this dependence, one can see that the limit on the NP effect in $a_e$ is so stringent~\cite{Hanneke:2008tm,Aoyama:2017uqe,Laporta:2017okg} that the NP effect in $a_\ell$ cannot be flavour blind, and must therefore break LFU, if one aims at explaining $a_\mu$\footnote{However, in this case, LFUV could originate from the SM Yukawa couplings like in the MSSM.}. 
	
	\vspace{2mm}
	{\bf CAA}:~The Cabbibo angle parametrizes the mixing among the first two generations of quarks and dominates the first row and first column CKM unitarity relations. Therefore, it can be used to check the consistency of different determinations of CKM elements within the SM and thus also to search for physics beyond the SM. Interestingly, a deficit in the first row and first column CKM unitarity is observed~\cite{Zyla:2020zbs}. This can be traced back to the fact that $V_{ud}$ extracted from super-allowed beta decays does not agree with $V_{us}$ determined from kaon and tau decays, when comparing (both) via CKM unitarity. The significance of these deviations crucially depends on the radiative corrections applied to beta decays~\cite{Marciano:2005ec,Seng:2018yzq,Seng:2018qru,Gorchtein:2018fxl,Czarnecki:2019mwq,Seng:2020wjq,Hayen:2020cxh,Hardy:2020qwl}, but also on the treatment of tensions between $K_{\ell 2}$ and $K_{\ell 3}$~\cite{Moulson:2017ive,Seng:2021nar} and tau decays~\cite{Amhis:2019ckw}. In the end $\sum_i\big|V_{ui}\big|^2= 0.9985(5)$ and $\sum_i\big|V_{id}\big|^2= 0.9970(18)$ should give a realistic representation of the current situation~\cite{Zyla:2020zbs}. Note that the fact that there is a deficit both in first row and first column CKM unitarity suggests that, if these tensions are due to NP, they should be related to $V_{ud}$ and thus beta decays. Interestingly, these deviations can also be interpreted as a sign of LFUV~\cite{Coutinho:2019aiy,Crivellin:2020lzu} since beta decays involve electrons while the most precise determination of $V_{us}$ comes from decays with final state muons. 
	
	\vspace{2mm}
	\begin{boldmath}$q\bar q\to e^+e^-$:\end{boldmath}~In the search for high-energetic oppositely charged lepton pairs, the CMS experiment observed 44 electron events with an invariant mass of more than $1.8\,$TeV, while only $29.2\pm3.6$ events were expected~\cite{CMS:2021ctt}, resulting in a $\approx4\sigma$ tension. As the number of observed muons is compatible with the SM prediction, this can be interpreted as a sign of LFUV and CMS provided the ratio of muons over electrons which reduces the theoretical uncertainties~\cite{Greljo:2017vvb}. Importantly, this CMS excess is compatible with the ATLAS limit~\cite{ATLAS:2020yat}, as also ATLAS observed slightly more electrons than expected.
	
	\vspace{2mm}
	\begin{boldmath}$\Delta A_{FB}$:\end{boldmath}
	This observable encodes the difference of the forward-backward asymmetry in $B\to D^*\mu\nu$ vs $B\to D^*e\nu$. Like for $R(K^{(*)})$ the muon and electron mass can both be neglected such that the form-factor dependence cancels and the SM prediction is, to the currently relevant precision, zero. Even though the corresponding measurements of the total branching ratios are consistent with the SM expectations~\cite{Glattauer:2015teq,Abdesselam:2017kjf}, recently Ref.~\cite{Bobeth:2021lya} unveiled a $\approx\!4\sigma$ tension in $\Delta A_{\mathrm{FB}}$, extracted from $B\to D^*\ell\bar \nu$ data of BELLE~\cite{Waheed:2018djm}.
	
	\vspace{2mm}
	\begin{boldmath}$\tau\to\mu\nu\nu$:\end{boldmath}
	Combining the ratios  $\tau  \to \mu,e \nu \bar \nu/\mu \to e \nu \bar \nu$ and $\tau  \to \mu \nu \bar \nu/\tau \to e \nu \bar \nu$, including the latest BELLE result~\cite{Belle:2013teo} and correlations~\cite{Amhis:2019ckw}, leads to a $\approx 2\sigma$ preference for constructive NP at the per-mille level in $\tau  \to \mu \nu \bar \nu$. On the theory side, QED corrections~\cite{Marciano:1988vm,Decker:1994ea} are relevant due to the high precision of the measurement.
	
	\section{Explanations}
	\label{explanations}
	
	Since the anomalies fit into the coherent patterns of LFUV beyond the SM, it is natural to ask how they could be explained in terms of new particles and new interactions. For a consistent renormalizable extension of the SM, only scalars bosons (spin 0), fermions (spin 1/2) and vectors bosons\footnote{Here a consistent extension requires some spontaneous symmetry breaking via a Higgs mechanism or some composite or extra-dimensional dynamics.} (spin 1) are at disposal. Here, we will consider four classes of SM extensions (focusing on heavy NP realized above the EW breaking scale):
	\vspace{-2mm}
	\begin{itemize}
		\item Leptoquarks (LQs): Scalar or vector particles that carry color and couple directly to a quark and a lepton~\cite{Buchmuller:1986zs,Dorsner:2016wpm}. They were first proposed in the context of the Pati-Salam model~\cite{Pati:1974yy}, Grand Unified Theories (GUTs)~\cite{Georgi:1974sy,Dimopoulos:1980hn,Senjanovic:1982ex,Frampton:1989fu,Witten:1985xc} and in the R-parity violating MSSM (see e.g. Ref.~\cite{Barbier:2004ez} for a review).
		\vspace{-2mm}
		\item $W^\prime$ bosons: Singly charged, QCD neutral vector particles. They appear as Kaluza Klein excitations of the SM $W$ in composite~\cite{Weinberg:1962hj,Susskind:1978ms} or extra-dimensional models~\cite{Randall:1999ee} as well as in models with additional $SU(2)$ factors, including left-right-symmetric models~\cite{Mohapatra:1974gc}.
		\vspace{-2mm}
		\item $Z^\prime$ bosons: Neutral (color and electric charge) vector bosons. They can be singlets under $SU(2)_L$ but also neutral components of an $SU(2)_L$ multiplet. Again, they can be resonances of the SM $Z$ or originate from an abelian symmetry like $B-L$~\cite{Pati:1974yy} or gauged flavour symmetries~\cite{Froggatt:1978nt,He:1991qd}.
		\vspace{-2mm}
		\item New scalars and fermions (S/F): In this category we pigeonhole all vector-like fermions as well as all scalar particles that are not LQs. Vector-like fermions appear in GUTs~\cite{Hewett:1988xc,Langacker:1980js,delAguila:1982fs}, composite models or models with extra dimensions~\cite{Antoniadis:1990ew,ArkaniHamed:1998kx} and vector-like leptons are involved in the type I~\cite{Minkowski:1977sc,Lee:1977tib} and type III~\cite{Foot:1988aq} seesaw mechanisms. New fermions and scalars could be supersymmetric partners of SM particles~\cite{Haber:1984rc} and the MSSM also includes additional Higgses like in the 2HDMs~\cite{Chanowitz:1985ug,Branco:2011iw}.
	\end{itemize}
	
	\begin{boldmath}{$b\to s\ell^+\ell^-$:}\end{boldmath}~As these processes are suppressed in the SM, the required $O(20\%)$ NP effect (w.r.t.~the SM) is small and we have three different classes of solutions:
	1)~A $Z’$ boson with a flavour violating couplings to bottom and strange quarks can account for the anomaly at tree-level~\cite{Buras:2013qja,Gauld:2013qba,Altmannshofer:2014cfa,Crivellin:2015mga,Crivellin:2015lwa,Niehoff:2015bfa}. Even though one in general expects an effect in $B_s-\bar B_s$ mixing~\cite{DiLuzio:2017fdq}, and the $Z^\prime$ can be produced resonantly at the LHC (see e.g.~\cite{Allanach:2015gkd}), such a solution is viable if the couplings to first generations quarks are suppressed and the models possesses an approximate global $U(2)$ flavour symmetry to protect it from $K^0-\bar K^0$ and $D^0-\bar D^0$ mixing~\cite{Calibbi:2019lvs}. 
	2)~Three LQ representations, the scalar triplet ($S_1$), vector singlet ($U_1$) and vector triplet ($S_3$), can already at tree-level give a good fit to $b\to s\ell^+\ell^-$ data~\cite{Hiller:2014yaa,Alonso:2015sja}. The couplings to electrons should be small because of $\mu\to e\gamma$~\cite{Crivellin:2017dsk} and $\mu\to e$ conversion. As effects in $b\to s\gamma$ and $B_s-\bar B_s$ mixing are only generated at the loop-level, these processes are not constraining and also the LHC bounds are weak~\cite{Diaz:2017lit} as no coupling to first generation quarks are needed.
	3)~Loop-effect involving box diagrams with new heavy scalars and fermions~\cite{Gripaios:2015gra,Arnan:2016cpy,Grinstein:2018fgb,Arnan:2019uhr} or top quarks~\cite{Aebischer:2015fzz} (either in combinations with LQs~\cite{Becirevic:2017jtw} or a $Z^\prime$~\cite{Kamenik:2017tnu}) can account for the anomaly if NP is at or below the TeV scale. 
	
	\begin{boldmath}{$b\to c\tau\nu$:}\end{boldmath}~As this transition is tree-level mediated in the SM, also a tree-level NP contribution is necessary to obtain the desired effect of 10\% w.r.t the SM (for heavy NP with perturbative couplings). As it is a charged current process, the only possibilities are  charged Higgses~\cite{Crivellin:2012ye,Fajfer:2012jt,Celis:2012dk}, $W’$~bosons~\cite{Bhattacharya:2014wla,Greljo:2015mma,Boucenna:2016qad,Greljo:2018ogz,Robinson:2018gza,Asadi:2018wea,Carena:2018cow} (with or without right-handed neutrinos) or LQs~\cite{Sakaki:2013bfa,Bauer:2015knc,Freytsis:2015qca,Fajfer:2015ycq}. The first two options are disfavoured by the $B_c$ lifetime~\cite{Celis:2016azn,Alonso:2016oyd} and/or LHC searches~\cite{Bhattacharya:2014wla,Greljo:2015mma}, leaving LQ as the best option for a full explanation. However, also in this case, a solution is not trivial, as constraints from $B_s-\bar B_s$ mixing, $B\to K^*\nu\nu$ and LHC searches must be avoided. Therefore, either the $SU(2)_L$ singlet vector LQ~\cite{Calibbi:2015kma,Barbieri:2016las,DiLuzio:2017vat,Calibbi:2017qbu,Bordone:2017bld,Blanke:2018sro,Crivellin:2018yvo} or the singlet-triplet model~\cite{Crivellin:2017zlb,Crivellin:2019dwb,Gherardi:2020qhc} that can avoid these constraints are particularly interesting. 
	
	\vspace{2mm}
	\begin{boldmath}$a_\mu$:\end{boldmath}~As the deviation from the SM prediction is of the order of its EW contribution, heavy NP around the TeV scale must posses an enhancement factor\footnote{See Ref.~\cite{Athron:2021iuf} for a recent overview on NP explanations of $a_\mu$.}. This can be provided via chiral enhancement meaning that the chirality flip does not originate from the muon Yukawa but from a larger coupling of NP to the SM Higgs doublet. In the MSSM, this factor is $\tan\beta$~\cite{Everett:2001tq,Feng:2001tr} and $a_\mu$ has been considered for many years as the smoking gun of the MSSM~\cite{Stockinger:2006zn}, however, due to the stringent LHC bounds minimal versions with universal scalar masses~\cite{Nilles:1983ge} cannot account for it anymore~\cite{Costa:2017gup} while the general MSSM still can. Alternatively, models with generic new scalars and fermions can explain $a_\mu$~\cite{Czarnecki:2001pv,Kannike:2011ng,Kowalska:2017iqv,Crivellin:2018qmi,Crivellin:2021rbq} and there are two scalar LQ representations that address $a_\mu$ via a $m_t/m_\mu$ enhancement~\cite{Djouadi:1989md,Davidson:1993qk,Couture:1995he,Chakraverty:2001yg,ColuccioLeskow:2016dox}.
	
	\vspace{2mm}
	{\bf CAA}:~A sub per-mille effect in the determination of $V_{ud}$ suffices to explain the CAA. In order to extract $V_{ud}$ from beta decays knowledge of the Fermi constant, most precisely measured in muon decay~\cite{Tishchenko:2012ie}, is needed. However, as the Fermi constant could subsume NP, we have the following possibilities~\cite{Crivellin:2021njn}: 1) direct (tree-level) modification of beta decays 2) direct (tree-level) modification of muon decay 3) modified $W$-$\mu$-$\nu$ coupling 4) modified $W$-$u$-$d$ coupling. Note than the effect of a modified $W$-$e$-$\nu$ coupling in $V_{ud}$ cancels~\cite{Coutinho:2019aiy,Crivellin:2020lzu}. Option 1) could be realized by a $W’$~\cite{Capdevila:2020rrl} or a LQ~\cite{Crivellin:2021egp}, however in the latter case stringent bounds from other flavour observables arise. Possibility 2) can be achieved by adding a singly charged $SU(2)_L$ singlet scalar~\cite{Crivellin:2020klg}, a $W^\prime$~\cite{Capdevila:2020rrl} or $Z^\prime$ vector boson with flavour violating couplings~\cite{Buras:2021btx}, while option 3) and 4) can be realized by vector-like leptons~\cite{Crivellin:2020ebi,Kirk:2020wdk,Alok:2020jod} and vector-like quarks~\cite{Belfatto:2019swo,Branco:2021vhs,Belfatto:2021jhf}, respectively. Note that vector-like quarks could also solve the tension between the different determinations of $V_{us}$ while vector like leptons have the potential to improve the global EW fit.
	
	\vspace{2mm}
	\begin{boldmath}{$q\bar q\!\to\! e^+e^-$:}\end{boldmath}~As CMS analyzes “non-resonant” electrons, meaning that they do not originate from the on-shell production of a new particle, NP must be heavier than the LHC energy scale. Furthermore, since the muon channel agrees with the SM expectations, constructive interference in the electron channel is required. This is possible with NP coupling to first generation quarks and electrons with $O(1)$ couplings and masses around 10~TeV. Therefore, a $Z^\prime$~boson or a LQ coupling to electrons and first generation quarks~\cite{Crivellin:2021egp} have the potential to explain this CMS measurement. However, taking into account the information on the various $q^2$ bins of the muon to electron ratio, one can see that NP is only preferred by $\approx 3\,\sigma$~\cite{Crivellin:2021rbf} meaning that the excess of $4\,\sigma$ cannot be fully explained. \footnote{{Note that even though here an effect in electrons is required, while $b\to s\ell^+\ell^-$ points towards NP mainly related to muons, this is not a priori a contradiction. It is well possible that NP couples electrons to first generation quarks (explaining $q\bar q\!\to\! e^+e^-$) and muons to second and third generation quarks (explaining $b\to s\ell^+\ell^-$).}}
	
	\vspace{2mm}
	\begin{boldmath}$\Delta A_{FB}$:\end{boldmath}~A good fit to data requires a non-zero Wilson coefficient of the tensor operators. Importantly, among the set of renormalizable models, only two scalar LQ can generate this operator at tree-level and only the $SU(2)_L$ singlet gives a good fit to data~\cite{Carvunis:2021dss}. However, even in this case, due to the constraints from other asymmetries, $\Delta A_{FB}$ cannot be fully explained, but the global fit to $b\to c\mu\nu$ and $b\to c e\nu$ data can be improved by more than $3\sigma$~\cite{Carvunis:2021dss}.
	
	\begin{table}[t]
		\begin{center}
		\begin{tabular}{ l | c  c c c}
			&statistics  & experiment & theory & NP explanation \\\hline
			$b \to s\ell \ell$ & $\star\star\star\star\star$ & $\star\star\star\star\star$ & $\star\star\star\star$ & $\star\star\star\star$ \\
			$b \to c\tau \nu $ & $\star\star\star$ & $\star\star\star$ & $\star\star\star\star$ & $\star\star\star$\\
			$a_\mu$ & $\star\star\star\star$ & $\star\star\star\star$ & $\star\star\star$ & $\star\star\star$\\
			CAA & $\star\star\star$ & $\star\star\star\star$ & $\star\star$ & $\star\star\star\star\star$\\
			$q\bar q\to e^+e^-$ & $\star\star\star$ & $\star\star\star$ & $\star\star\star\star\star$ & $\star\star\star$\\
			$\Delta A_{FB}$ & $\star\star\star\star$ & $\star\star$ & $\star\star\star\star\star$ & $\star\star$\\
			$\tau\to\mu\nu\nu$ & $\star\star$ & $\star\star\star$  & $\star\star\star\star$ & $\star\star\star\star$
		\end{tabular}
	\end{center}
		\caption{{Quantitative} comparison, using a one to five stars rating, of the different anomalies pointing towards LFUV.}
		\label{table:Anomalies}
	\end{table}

	\vspace{2mm}
	\begin{boldmath}$\tau\to\mu\nu\nu$:\end{boldmath}~Explanations of $\tau\to\mu\nu\nu$ are very similar to NP explanations of the CAA via a modified Fermi constant (with $\tau\to\mu\nu\nu$ taking the role of $\mu\to e\nu\nu$). It can be achieved by a tree-level effect via a singly charged $SU(2)_L$ singlet scalar~\cite{Crivellin:2020klg}, a $W^\prime$ or a flavour violating $Z^\prime$~\cite{Buras:2021btx}. Alternatively, a modification of the $W$-$\tau$-$\nu$ coupling  through the mixing of vector like leptons or a $W^\prime$ boson is possible. Furthermore, a $Z^\prime$ coupling to muons and tau leptons can generate the right effect via box diagrams~\cite{Altmannshofer:2014cfa,Crivellin:2020oup}.
	
	\begin{figure}[t]
				\begin{centering}
		\includegraphics[width=0.7\textwidth]{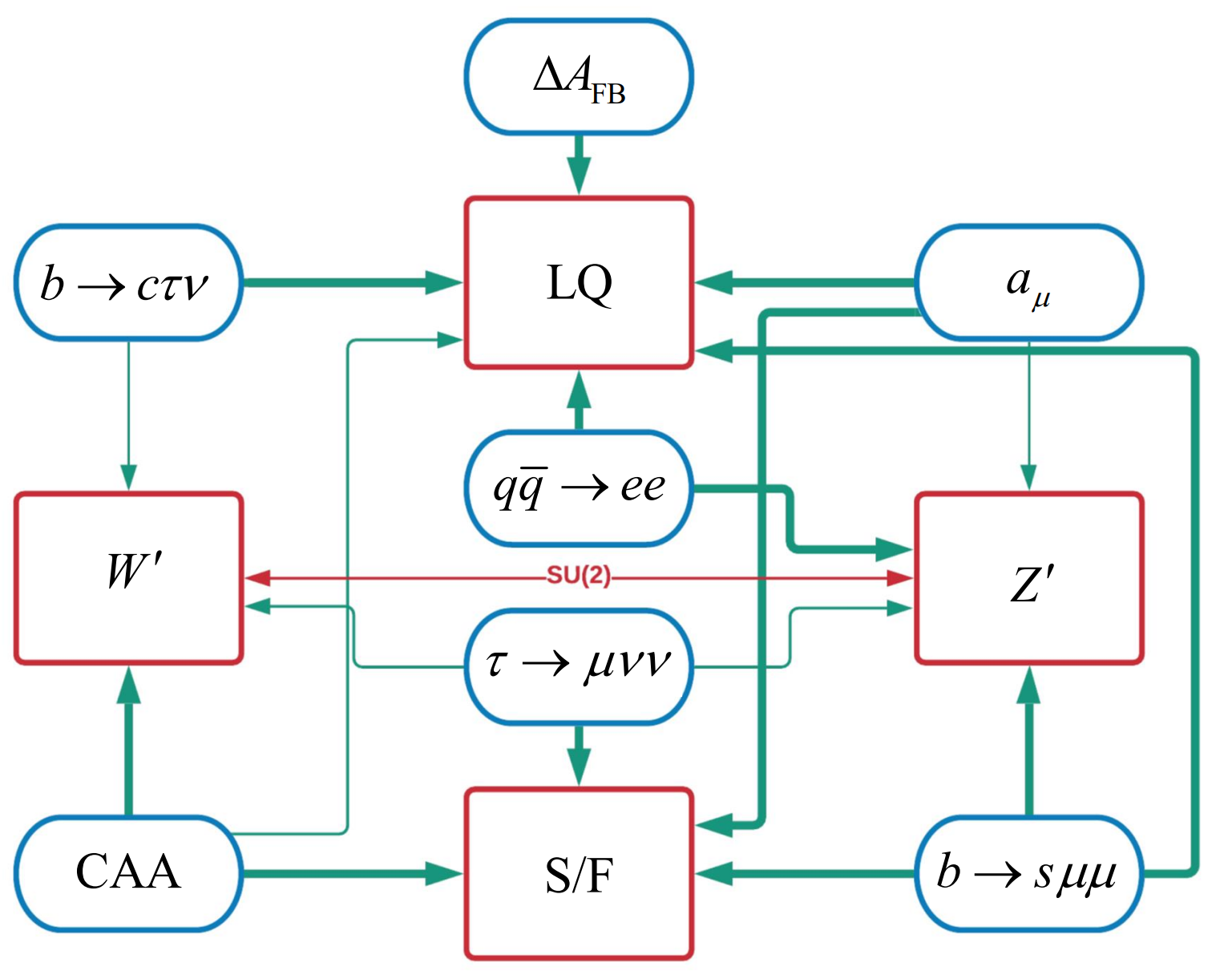}
		\caption{{Synthesis} of possible explanations (red boxes) of the anomalies (blue boxes). The arrows indicate to which extensions of the SM the anomalies point. Thick arrows stand for probable explanations without significant experimental or theoretical shortcomings while the thin ones indicate that the new particles can only partially explain the measurement or generate problems in other observables. The red arrow indicates that the $Z^\prime$ and $W^\prime$ could be components of a single $SU(2)_L$ triplet.}
		\label{fig:Explanations}
				\end{centering}
	\end{figure}
	
	\section{Conclusions and Outlook}
	\label{conclusions}
	
	In this letter we {proposed} that not only the ratios $R(K^{(*)})$, $R(D^{(*)})$ and $Q_5$ could be manifestations of LFUV physics beyond the SM but that also several other anomalies admit an interpretation in this context. We {display} these observables in Table~\ref{table:Anomalies} where we compare them regarding  four categories:
	\vspace{-1mm}
	\begin{itemize}
		\item Statistics: Statistical significance of the deviations from the SM prediction. 
		\vspace{-1mm}
		\item Experiment: Variety and dis-correlation of the available measurements. Robustness regarding systematic uncertainties.
		\vspace{-1mm}
		\item Theory: Solidity and coherence of the SM prediction.
		\vspace{-1mm}
		\item NP explanations: Explainability in terms of new particles and interactions, taking into account possible conflicts with other observables.
	\end{itemize}
	\vspace{-2mm}
	with a star rating from one to five. From this table we can see that $b\to s\ell^+\ell^-$ has top ratings for statistics and experiment and and also the theory prediction is very solid due to recent progress in the calculation of long distance charm computations~\cite{Gubernari:2020eft}. For $a_\mu$ the two weaker points are the SM prediction in regard to HVP and that the required NP effect is large, such that it is not easy to obtain in a UV complete model. $\Delta A_{FB}$ has only two stars for the NP explanation as the tension can only be eased but not fully explained. Both the CAA and $\tau\to\mu\nu\nu$ are very easy to explain via NP as only a per-mille effect is required but have weaknesses in their SM predictions and statistics, respectively.  
	
	As {we have shown that all} the anomalies fit into the coherent pattern of LFUV, common explanations are not only possible but even probably. We {synthesize} the {viable} SM extensions (LQs, S/F, $W^\prime$ and $Z^\prime$) which can account for them in Fig.~\ref{fig:Explanations}. From there we can see that the various anomalies point towards different possible extensions of the SM and that an extension by a single new field cannot explain all anomalies, however, in particular LQs {are} very promising solution. On the one hand, this could indicate that not all anomalies might be confirmed in the future. On the other hand, it is of course possible (and maybe even likely) that the SM is superseded by a more complicated theory with a sizeable number of new degrees of freedom. Importantly, even in case only one of the anomalies were confirmed, this would prove the existence of physics beyond the SM at the (multi) TeV scale which would not only constitute the biggest breakthrough in particle physics within the last decades but also provide a convincing physics case for future colliders {guiding the field into a new era}.  
	
	\section*{Acknowledgments}
		A.C. gratefully acknowledges the support by the Swiss National Science Foundation under Project No.\ PP00P21\_76884. JM acknowledges  financial support from the Spanish Ministry of Science, Innovation and Universities (PID2020-112965GB-I00/AEI/ 10.13039/501100011033) and  by ICREA under the ICREA Academia programme.

\end{document}